\documentclass[manuscript]{aastex631}

\submitjournal{ApJ}

\shorttitle{Radio Occultation using MOM}
\shortauthors{Aggarwal et al.}

\begin{document}

\title{Insights into Solar Wind Flow Speeds from the Coronal Radio occultation Experiment: Findings from the Indian Mars Orbiter Mission}

\correspondingauthor{Keshav Aggarwal}
\email{keshavagg1098@gmail.com}

\author[0000-0002-7004-8670]{Keshav Aggarwal}
\affiliation{Department of Astronomy, Astrophysics and Space Engineering (DAASE), \\
Indian Institute of Technology Indore,Indore, Madhya Pradesh, 453552}

\author[0000-0002-1276-0088]{R. K. Choudhary}
\affiliation{Space Physics Laboratory (SPL), Indian Space Research Organization, \\
Vikram Sarabhai Space Centre, Thiruvananthapuram, Kerala 695022, India}

\author[0000-0002-5333-1095]{Abhirup Datta}
\affiliation{Department of Astronomy, Astrophysics and Space Engineering (DAASE), \\
Indian Institute of Technology Indore,Indore, Madhya Pradesh, 453552}

\author{Roopa M. V.}
\affiliation{ISRO Telemetry Tracking and Command Network (ISTRAC), Bengaluru, Karnataka 560058, India}

\author{Bijoy K. Dai}
\affiliation{ISRO Telemetry Tracking and Command Network (ISTRAC), Bengaluru, Karnataka 560058, India}

\begin{abstract}

Using data collected by the Indian Mars Orbiter Mission in October 2021, we investigated coronal regions of the Sun by analyzing the Doppler spectral width of radio signals to estimate solar wind velocity. A simplified equation is introduced to directly relate these two parameters. The study focuses on observations conducted from October 2 to October 14, 2021; a relatively quiet phase of solar cycle 25. The analysis targeted the coronal region within heliocentric distances of 5–8 $R_\odot$,  near the ecliptic plane. In this region, solar wind velocities ranged from 100 to 150 km/s, while electron densities were on the order of 10$^{10}$ m$^{-3}$. We also compared our results with electron density observations and models derived from previous studies. Though the decrease in the electron densities with respect to increasing helio-centric distance matches quite well with the theoretical models, MOM estimates fall at the lower edge of the distribution. This difference may be attributed to the prolonged weak solar activity during the MOM observations, in contrast to prior studies conducted during periods of comparatively higher solar activity in earlier solar cycles.

\end{abstract}

\keywords{Solar wind (1534), Radio occultation (1351), Solar corona (1483)}

\section{Introduction} \label{sec:intro}
The Solar wind originates from the Sun as a stream of charged particles, consisting of ions and electrons, which escape into the interplanetary medium. The escape of particles is due to the extreme temperatures of coronal plasma reaching millions of degrees, resulting in tremendous thermal agitation speed of the plasma particles. The solar wind accelerates from subsonic to supersonic speeds in the middle and upper corona, which stretches up to a few solar radii from the Sun's surface \citep{West2023}. The acceleration mechanisms of the solar wind have been observed using various techniques, including radio occultation, which suggests that coronal heating mechanisms play a vital role in solar wind acceleration, with the acceleration region identified between 2 and 10 $R_{\odot}$ \citep{West2023, Jain_2023, Jain_2024}. The solar wind is a major driver in solar-terrestrial interactions and affects Earth's magnetosphere and ionosphere \citep{Jain_2024b}. Studies indicate that Alfvén waves, through their interactions with solar wind particles, contribute to the outward acceleration \citep{Rivera2024}. It is generally classified into three main categories: steady fast winds in coronal holes, unsteady slow winds created by temporarily open streamer belts, and transient winds originating from large coronal mass ejections \citep{Marsch1999}. Slow winds come from the lower latitude regions with velocities $\sim 200-300 $ km/s. The transient winds are the Coronal Mass Ejections (CMEs) which are large clouds of plasma ejected from the Sun and can travel with speeds up to as high as 1000 km/s \citep{Wang1990, Wang1991, Sakao2007, McComas2008, Suzuki2012}. Fast solar winds, on the other hand, have a speed of 700-800 km/s at Earth orbit, are more steady, and stream from coronal holes which correspond to open magnetic flux tube regions. CMEs and fast solar wind have the potential to cause geomagnetic storms that interfere with power grids, satellite operations, and navigation systems \citep{Tsyganenko2014, Marov2015, Hands2018, Berger2023, Parker2024}.

The measurement of Solar Winds began in 1960 using 3-electrode charged particle traps on the early space missions led by the Soviet Space Agency \citep{Gringauz1960}. Observations of the corona and solar wind thereafter have been conducted both through the in situ and remote sensing techniques. In-situ observations involve sending probes in the interplanetary medium, while remote sensing involves observing signals naturally generated within the coronal region or conducting studies using probe signals from natural or man-made radio sources. There have been several missions to the Sun that have carried out in-situ measurements of the solar wind, with the most recent being the Parker Solar Probe (PSP) launched by NASA in 2018 and the Aditya-L1 mission launched by ISRO in September 2023 \citep{Tripathi2023}. The Parker Solar Probe is equipped with instruments like the SWEAP and FIELDS \citep{Bale2016, Kasper2016}. Other probes to the Sun include the Solar Orbiter launched by ESA in 2020, as well as the Helios Probes launched in 1974 and 1976, which traveled as close as 60 solar radii and made measurements of the solar wind plasma and changes in the solar magnetic field \citep{Roberts1987, Muller2020, Marirrodriga2021}. Additionally, there was a sample return mission named Genesis launched by NASA in 2001, which retrieved samples from the Sun-Earth L1 point, a prime vantage point for studying the Sun \citep{Wiens2021}. However, in-situ observations are challenging due to the extreme conditions in coronal plasma, such as extremely high temperatures and very low electron densities, and only yield information about a single point in space.

\begin{figure}
\plotone{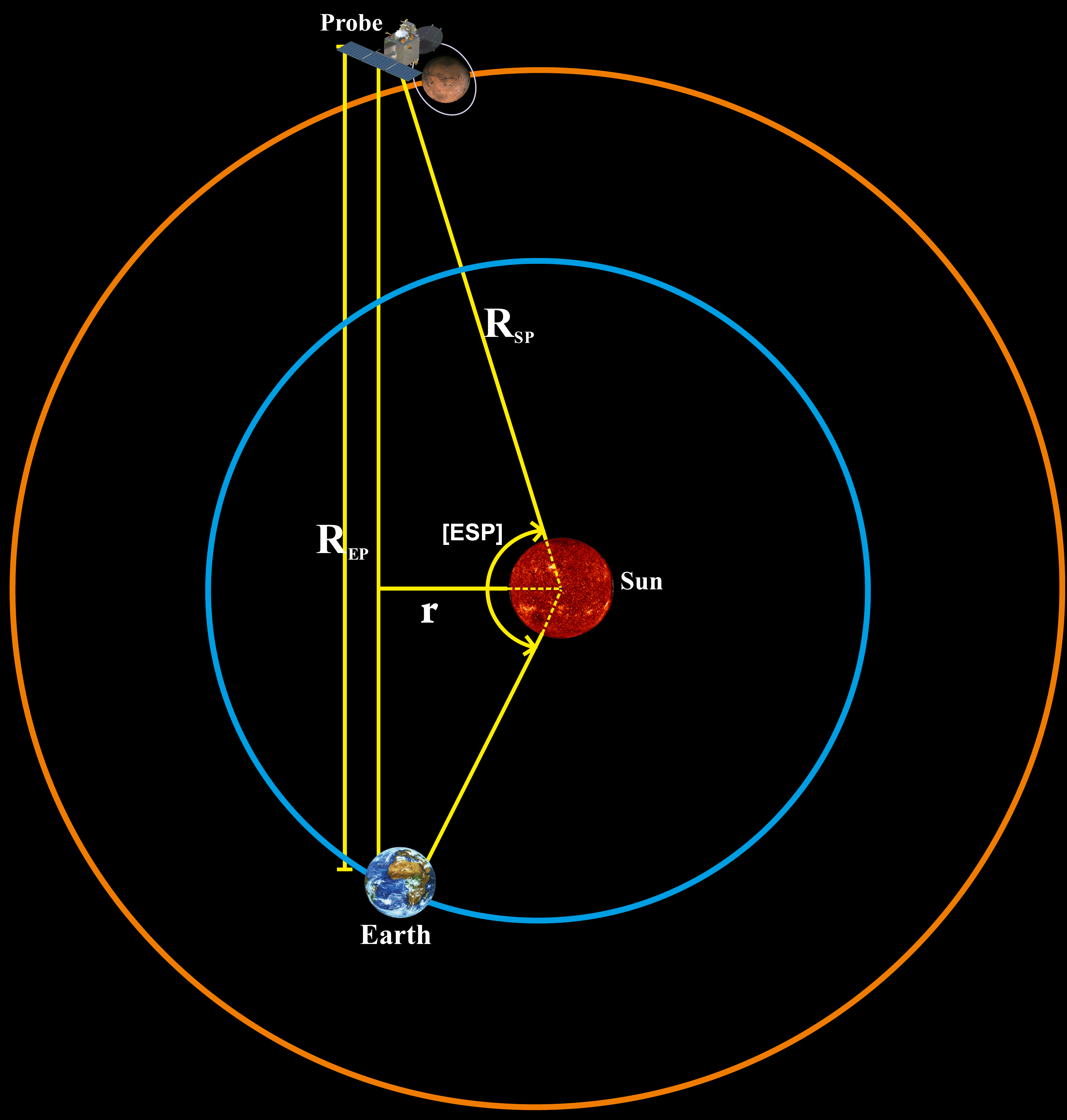}
\caption{Graphical representation of the geometry during a radio occultation experiment. \label{fig:experiments}}
\end{figure}

Among the active remote sensing techniques, Radio Occultation (RO) is a well-established method for probing the solar corona and studying the behavior of the solar wind, and observations conducted by various spacecraft have been crucial in studying the Sun's activity over multiple solar cycles. Notably, Pioneer 6 \citep{Woo1976}; HELIOS (1975–1976, \cite{Paetzold1987, Wexler2019}); Pioneer 10/11 (1978, \cite{Woo1978, Coles1991}); Viking (1979, \cite{Tyler1977}); Ulysses (1991, \cite{Efimov2005}); Galileo (1994, 2000, \cite{Wohlmuth2001}); Nozomi (2000-2001, \cite{Imamura2005, Tokumaru2012}); Cassini (2002, \cite{Morabito2007}); MESSENGER (2009, \cite{Wexler2019}); Rosetta, MEX, and VEX (2010, \cite{Paetzold1996}); Akatsuki (2011, \cite{Miyamoto2014, Ando2015, Wexler2020, Jain_2023, Jain_2024}); MAVEN (2014, \cite{Withers2018, Withers2020}); and the Indian Mars Orbiter Mission (MOM) (2015, \cite{Jain2022, Jain_2024b}) have contributed to these investigations. During these experiments, a spacecraft’s radio signal passes through the solar corona when it is occulted by the Sun as seen from Earth, allowing valuable tests to be conducted as the satellite signal traverses the near-sun medium at very close proximity to the Sun. The irregularities in plasma density within the solar plasma act as scattering agents for radio waves as they travel through the solar plasma, providing crucial insights into the solar wind’s velocity, density, and magnetic field characteristics \citep{Woo1978, Woo1979, Bird1982, Bird1990, Patzold2005, Efimov2015, Jain2022}. This enables observation of the specific region around the Sun while posing no risk to the spacecraft and since no additional instruments are required for these experiments, they are very cost-effective as well.  However, the RO technique is unable to detect fine-scale spatial details in the Solar Corona, and the coverage is also limited to particular observational geometries. Additionally, Doppler due to factors other than the medium of interest, like interference, noise, and data handling techniques also affect the results, which are removable using different techniques to a reasonable extent \citep{Tripathi2022}. However, this technique can be used to infer solar wind parameters from the unreachable region of the Sun \citep{Jain_2023}.

In the solar RO experiments, the solar offset distance or the proximal point ($r$), which is the closest point of approach between the signal path and the Sun's center, is a critical parameter for measuring the impact of solar wind on radio waves, as it determines how deeply the signal penetrates the solar corona. The geometry of the experiment is illustrated in Figure \ref{fig:experiments}, which highlights how the spacecraft's radio signal propagates through the corona at varying radial distances $r$ from the Sun, measured in Solar radii ($R_{\odot}$). Here, $R_{EP}$ is the distance of the probe from the Earth in AU (1 AU = $1.496 \times 10^{11}$ m), $R_{SP}$ is the distance between the Sun and the probe in AU, and $[ESP]$ is the Earth-Sun-Probe angle measured in radians.

\begin{figure*}
\plotone{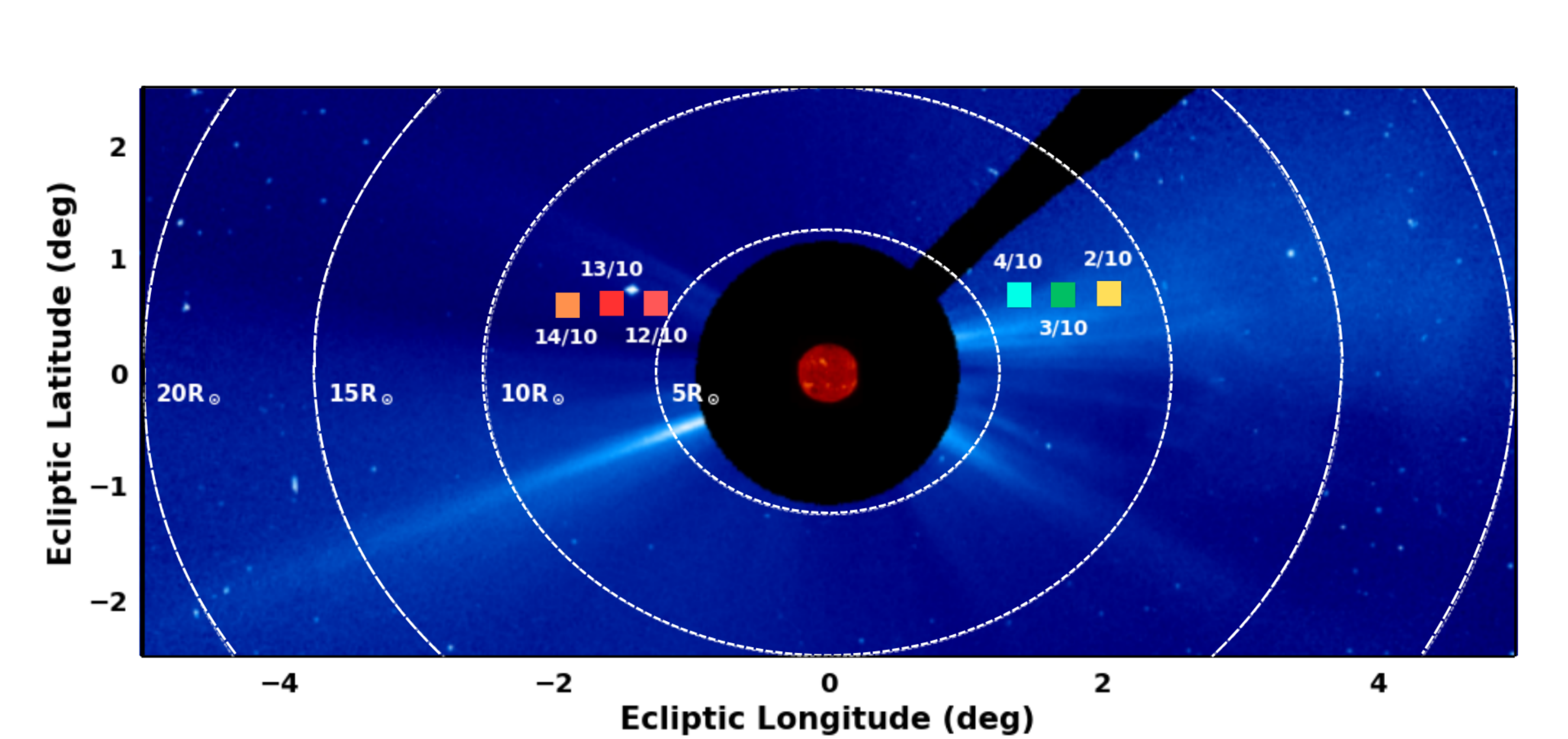}
\caption{This graphic displays the position of MOM (marked by colored boxes) in its orbit around Mars in relation to the Sun (composite image made using SDO-AIA 171 and SOHO-LASCO C2/C3 white light images), as observed from Earth on dates of experiment. The specific dates of observation are identified by the numbers near each colored box.}
\label{fig:transit}
\end{figure*}

The region under investigation is critical, as it contributes valuable observational data on the solar wind acceleration zone within the closest range that will be measured in situ by the Parker Solar Probe. While the dynamics of the lower corona are dominated by the magnetic field and have been extensively studied with soft X-ray/EUV imagers like Hinode/XRT, PROBA2/SWAP and the SDO/AIA, in the extended corona solar wind outflows dominate the dynamics and the main method of study is the use of white-light occulted coronagraphs like SOHO/LASCO and STEREO \citep{Nitta2021}. The transition between these two regimes is currently not well understood due in part to the relative paucity of measurements available in the mid-coronal gap, and the lack of a uniform measurement technique applicable across the adjacent regions. The current study intends to contribute to the study of this region by tracking the radio communications from India's Mars Orbiter Mission (MOM) through the solar corona, deriving solar wind speeds, and studying the evolution of solar wind as it crossed various heliocentric distances between 5 - 10 Solar Radii.

We used the S-band (2.3 GHz) RO experiments, conducted by the Indian MOM spacecraft from October 2 to October 14, 2021, for measurements of Solar wind velocities in the middle and outer coronal regions. This is significant for multiple reasons: Firstly, the period of observation of this study was a relatively quiet period, which gave us an opportunity to track the evolution of solar wind speeds from the source to different heliocentric distances. Secondly, the orbital inclination of the spacecraft during the period of this experiment enabled us to investigate the solar corona at close distances reaching as near as $5-8 R_{\odot}$ from the Sun in the equatorial plane. Fig \ref{fig:transit} gives a graphic display of the position of MOM in relation to the Sun, as observed from the Earth, while in its orbit around Mars. The specific dates of observation are identified by the numbers near each color box. Contour lines are drawn at a distance of 5 $R_{\odot}$ each to give an idea of the position of the probe.

In this manuscript, section \ref{sec:maths} discusses the methodology applied in this study, explaining the experiment, calculation of parameters like Doppler shift, TEC, plasma frequency, and angular broadening. Finally, the velocity of solar wind is calculated in section \ref{sec:observations}, which is then compared with the values of solar wind velocity measured in previous experiments, and the conclusions are given in section \ref{sec:ANSWER}.

\begin{deluxetable*}{lcc}
\tablenum{1}
\tablecaption{MOM Mission Parameters\label{tab:mission_parameters}}
\tablewidth{0pt}
\tablehead{
\colhead{Parameter} & \colhead{Value}
}
\startdata
\multicolumn{3}{c}{Mission Parameters} \\
\hline
Launch Date & 5th November 2013 & \\
Martian Orbit insertion & 24th September 2014 & \\
Planned mission duration & 6 months & \\
Duration of operations & 8 years, 9 days & \\
Apoapsis & $\sim$ 72000 km & \\
Periapsis & $260$ to $\sim 550$ km & \\
Orbital period & $\sim$ 66h & \\
Inclination & $\sim 150^{\circ}$ & \\
Velocity & $\sim 4.5 km/s$  near periapsis \\
\hline
\multicolumn{3}{c}{Antenna Parameters} \\
\hline
Diameter & 2.2m  HGA \\
Operating Frequency & 2292.96 MHz & \\
Power requirement & 440 W  DC power \\
Beamwidth & $\pm 2^{\circ}$  Right Circularly Polarized \\
Peak gain & 31dB & \\
\enddata
\end{deluxetable*}

\newpage
\section{Methodology}
\label{sec:maths} 

Indian Mars Orbiter Mission (MOM), undertook a successful journey of 300 days to enter Martian orbit and went on to perform a lot more than the intended mission lifespan of 6 months and marked the achievement of 8 years of continuous operations on September 24, 2022, \citep{Arunan2015, Bhardwaj2016, Jain2022}. Solar occultation experiments were conducted during the October 2021 Mars Solar conjunction using the MOM spacecraft which was in an elliptical orbit around Mars, with relevant parameters given in table \ref{tab:mission_parameters}. The 2.2 m high-gain parabolic offset disc reflector antenna of the MOM spacecraft, with oscillators having an Allan variance of the order of $ 10^{-11}$ was used for communications with Earth which transmitted and received telemetry, tracking, and commands data between the probe and the Indian Deep Space Network receiving station \citep{Ramamurthy2015}. The S-band (2.29 GHz) signals from the spacecraft were received by the high-gain 32-meter antenna stationed at the Indian Deep Space Network (IDSN) facility situated in Bangalore using a radio science receiver. Details on the open-loop radio science receivers can be found elsewhere \citep{bedrossian2019}. As S-band signals are transparent to the Earth’s atmospheric window they can propagate through the Earth's atmosphere without getting reflected by Earth’s atmosphere and hence experience minimal attenuation, absorption, or scattering making the frequency a good choice for conducting this experiment.

As mentioned above, when the experiments were conducted, the Sun was relatively quiet during the ascending phase of the solar cycle 25. The unique geometry of the probe and its highly elliptical orbit around the planet Mars allowed us to probe the solar corona at a very close range from $5-8R_{\odot}$, close to the ecliptic plane as shown in Fig \ref{fig:transit}. The characteristics that constitute a radio signal, such as its intensity (amplitude), phase or frequency, and polarisation, change as the signal travels through an ionized plasma medium and effects like amplitude scintillation, phase fluctuations, frequency Doppler shifts, signal broadening, and Faraday rotation are observed which happen due to the interaction of electromagnetic radio waves with the free electrons and ions in the plasma.

Analyzing these radio signals provides information about the turbulence level, coronal electron density spectrum, estimates of solar wind speeds in the inner heliosphere and magnetic configuration in coronal areas \citep{Jain2022, Jain_2023, Jain_2024}. The received signals are recorded in the open-loop mode in which the radio science receiver (RSR) utilizes a superheterodyne to translate the received signal from its original frequency to an intermediate frequency (IF), allowing for easier processing and filtering within a specific frequency range. In this process, a nonlinear device mixes the received signal with a local oscillator signal which creates two new frequencies: the sum and the difference of the original frequencies. A low-pass filter isolates the difference frequency, which becomes the downconverted signal. These signals were transformed into channel-level data and stored in a raw data exchange format (RDEF) with a sampling rate of 100 KHz per second. More details can be found in the CCSDS (Consultative Committee for Space Data Systems) \href{https://public.ccsds.org/Pubs/506x1b1.pdf}{blue book}. The amplitude of the received complex signal is then obtained by combining the I and Q values in the I+iQ form from the raw data and reading them into the integer data format. We then pad these values with zeroes such that the length becomes equal to $2^{19}$, i.e., a power of two that improves efficiency, reduces artifacts such as spectral leakage, enhances frequency resolution, and ensures compatibility with optimized FFT implementations. After applying FFT (fast Fourier transform) to the complex signal, frequency bins are generated, converting the signal power information from the time domain to the frequency domain. The square of the magnitude of the FFT of the received signal gives the power spectrum, which we assume to be a Gaussian of the form defined by the following equation:

\vspace{-0.5cm}

\begin{equation}
    S(\omega) = \frac{P}{(2 \pi B_S^2)^{1/2}} exp \left( \frac{-(\omega - \Omega)^2}{2B_S^2} \right)
	\label{eq:gaussian}
\end{equation}

\begin{figure}[htb]
\centering
 \includegraphics[scale=0.45]{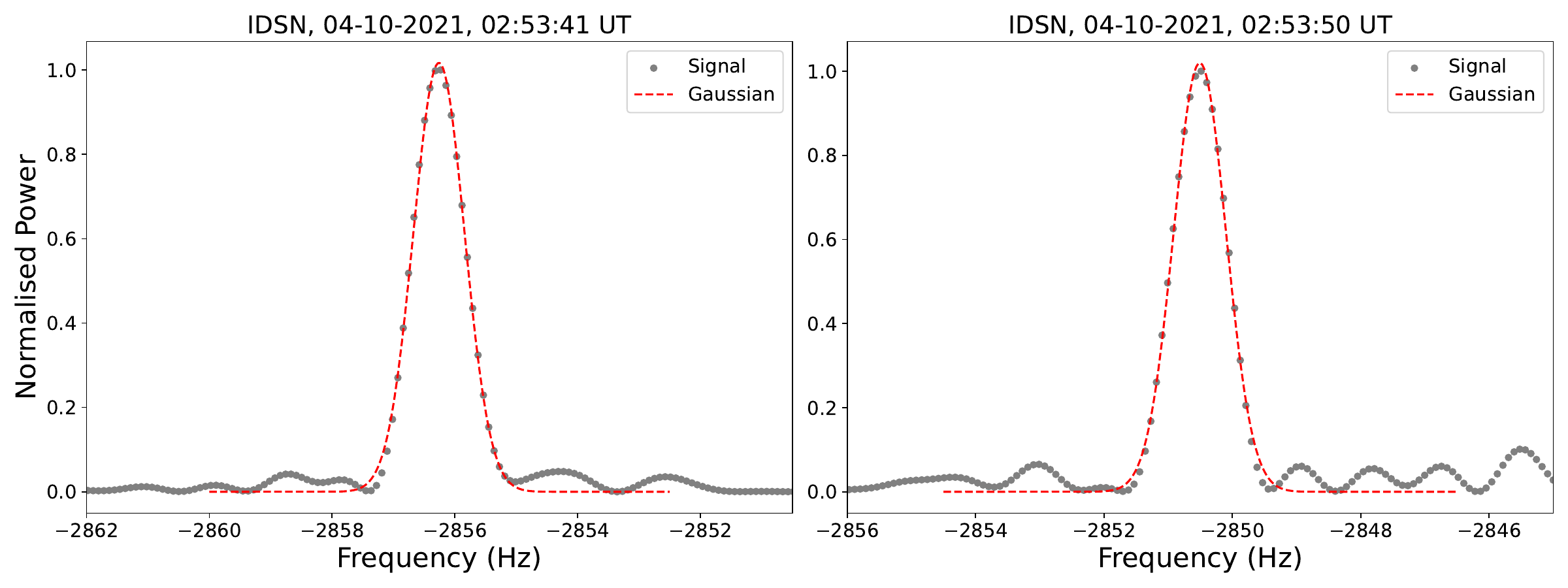}
 \caption{Gaussian fit to the spectrogram of the received signal for a 1-s data frame.}
\label{fig:spectrum}
\end{figure}

The zeroth moment of $S(\omega)$, representing the total power of the signal corresponds to the norm of the distribution and is denoted by $P$ in Equation \ref{eq:gaussian}. The first moment, which represents the Doppler velocity of the received signal, is denoted by $\Omega$ in Equation \ref{eq:gaussian}. The second moment, corresponding to the variance, describes the spectral width and is denoted as $B_S$ in Equation \ref{eq:gaussian}. Spectral width quantifies the dispersion of the signal over Doppler velocities, often referred to as spectral broadening, which arises from turbulence and the presence of particles with varying terminal velocities in the sampled volume. These three parameters—total power, frequency shift, and spectral width—are fundamental to characterizing the Doppler spectrum \citep{Woodman1985} and are estimated in occultation experiments \citep{Tripathi2022b, Tripathi2022, Jain2022, Jain_2023, Jain_2024}. Since $S(\omega)$ cannot be directly obtained in practice, statistical estimates are derived at a finite discrete set of N frequencies, $S(\omega_i)$. The following estimators are employed to compute these three parameters:

\begin{equation}
    P = \sum_{i=1}^N S'(\omega_i)
	\label{eq:norm}
\end{equation}

\begin{equation}
    \Omega = \frac{1}{P} \sum_{i=1}^N \omega_i S'(\omega_i)
	\label{eq:mean}
\end{equation}

\begin{equation}
    B_S^2 = \frac{1}{P} \sum_{i=1}^N (\omega_i - \Omega')^2 S'(\omega_i)
	\label{eq:variance}
\end{equation}

\begin{figure} [htb]
 \includegraphics[scale=0.4]{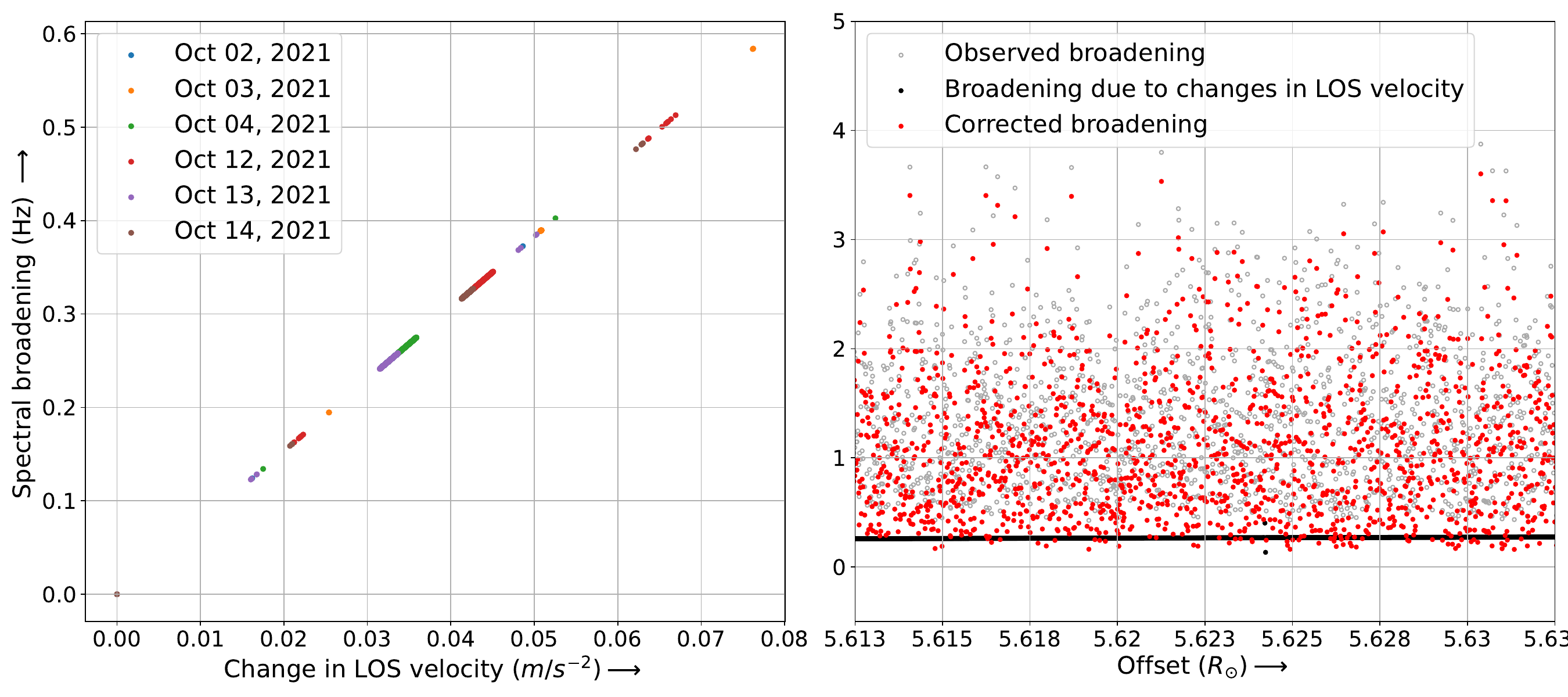}
\caption{Left panel : Doppler broadening due to rate of change of LOS Doppler velocity at varying helio-centric distances. Right panel : Change in the doppler broadening after removing the broadening due to the changes in LOS velocity. \label{fig:LOS_mitigation}}
\end{figure}

The spectral broadening of a signal (second moment) measures the increase of the bandwidth at the full-width-half-maximum (FWHM). While it is dependent on both electron density fluctuations and solar wind bulk plasma fluctuations, it is independent of the SEP angle, which makes it a useful tool to probe transients and features closer to the Sun \citep{Woo1976, Woo1977, Woo1978}. Fig \ref{fig:spectrum} shows Gaussian fits to a few samples of spectrogram of the received signal for a 1-s data frame at 02:53:41 UT (left panel), and 02:53:50 UT (right panel). The gray color points in the Figure represent the spectral power density while the dashed-dotted lines in red represent Gaussian fits.

There can be several other factors that can contribute to the broadening in the peak of the Doppler spectra, which include fluctuations in the intervening medium as well as the changes in the line-of-sight (LOS) Doppler velocity (relative velocity between the probe and the ground station along the line of sight) \citep{Bird1982, Tripathi2022}. The Doppler broadening in the received signal due to the changes in LOS velocity can be estimated using a method as described in \cite{Tripathi2022}:

\begin{equation}
   \Delta B_s = \frac{\left| d(V_{gs}) \right|}{c} \times f
    \label{eq:LOS_doppler}
\end{equation}

where $d(V_{gs})$ is the change in the LOS velocity, $f$ is the signal frequency in Hz, and $c$ is the speed of light \citep{Tripathi2022}. The left panel of fig \ref{fig:LOS_mitigation} shows how the value of spectral broadening changes before and after accounting for this error. In this figure, grey circles represent observed Doppler broadening while the points in red represent the same after removing contributions due to LOS changes. The right panel of figure \ref{fig:LOS_mitigation} shows how the rate of change of LOS Doppler velocity is related to the broadening in the Doppler spectrum. Since the rate of change in the LOS Doppler velocity is less, its impact on the spectral broadening is also minimal.

\begin{figure} [htb]
 \includegraphics[scale=0.4]{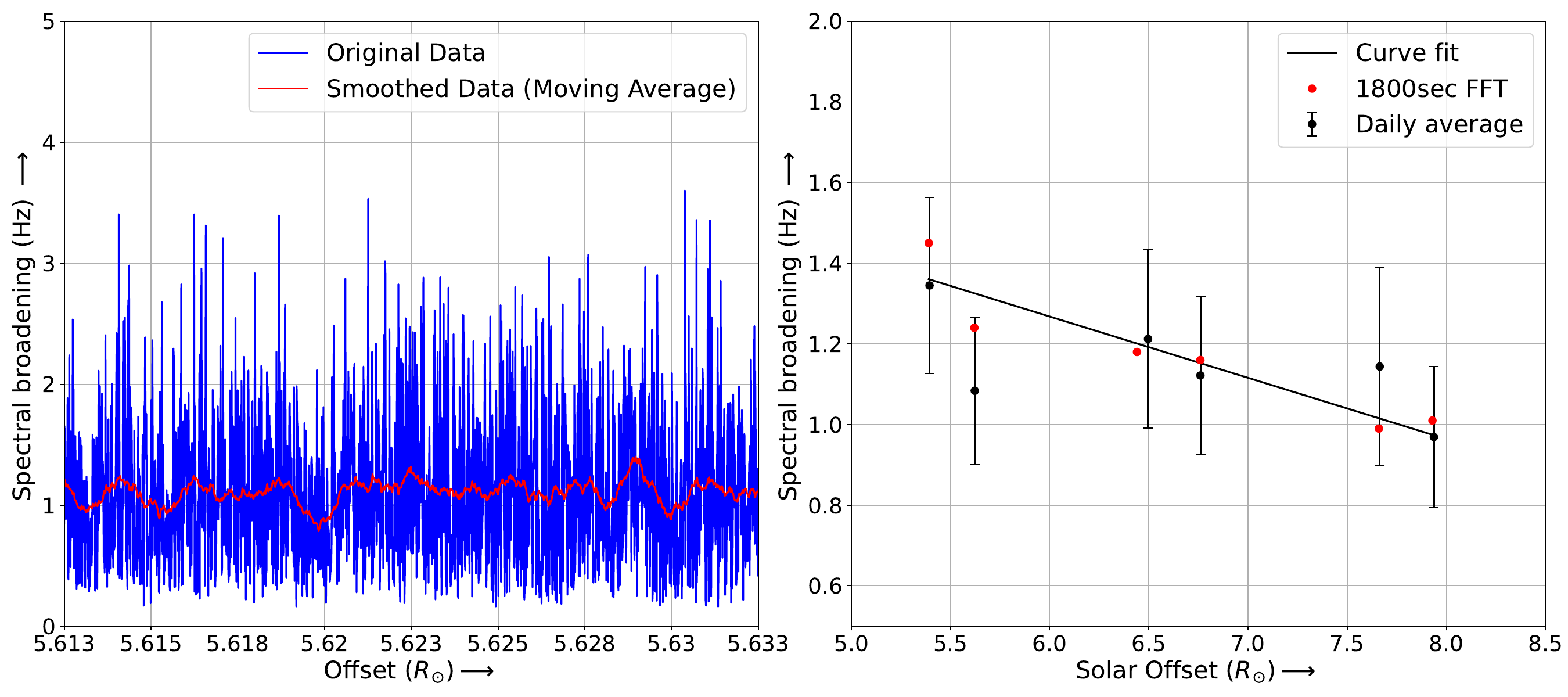}
\caption{Left panel : Computed values of second moment compared against the moving average for a single day of observations. The original data is shown in blue, while the smoothed data after removing extreme outliers is in red. Right panel : Daily average values of the second moments compared against each other, as a function of increasing solar distance.}
\label{fig:width}
\end{figure}

Figure \ref{fig:LOS_mitigation} shows variations in the Doppler broadening in the 1 second sample data observed on 04 Dec. 2021. Though the Doppler broadening has significant fluctuations in the values, the values are of the order as seen during previous missions using S-band of radio signals \citep{Woo1976, Woo1977, Woo1978, Woo1979, Ho2002, Morabito2003}. To minimize the fluctuations, we used a 1-minute moving average, as illustrated in the left panel of Figure \ref{fig:width}, assuming that the actual solar offset distances do not change too much during the average period. The smoothed data provides a better estimate for ambient Doppler broadening in the radio signals while they traverse through varying coronal regions. We compared the daily averages across the observation period, which show that the Doppler broadening decreases with increasing distance from the Sun, as demonstrated in the right panel of Figure \ref{fig:width}, and have plotted the standard deviation to illustrate the variability within individual measurements. In addition, we analyzed the daily averages by applying the power spectrum to the entire 1800-second data segment and determined the effective spectral width across the observation period. The daily estimated values of the average spectral width is also shown as a red color point in the right panel of Figure \ref{fig:width}. We may note that both the techniques yeild almost similar values for the average daily spectral width while aligning well with the slopes observed in previous studies by \cite{Woo1978} and \cite{Ho2002}. The Allan variance of the oscillator is $10^{-11}$, and the error due to the instability in the generated signal for the frequency is $10^{-11}\times f = 0.0229$ Hz \citep{Tripathi2022b}. The sum of the standard deviation in the daily averages and the error due to instability in the received signal gives us the final value of error in the estimation of the spectral broadening of the signal.

In this present work, we use these spectral broadening values to derive electron densities and solar wind speeds in the middle-upper coronal region as explained in the next section.

\section{Results and Discussion}
\label{sec:observations}

\subsection{Estimation of Electron density fluctuations}

If the signal is propagating through a turbulent medium that induces random phase changes at smaller time scales than the FFT integration time, it is believed that this would lead to a broadening of the centerline in the FFT \citep{Lipa1979}. Among the various reasons for broadening the received signals, an important one is the variations in the electron density of the coronal medium through which the signal travels \citep{Bird1982}. As it depends on both the electron density fluctuations and solar wind velocity, spectral broadening becomes a useful measure for probing the solar wind as well \citep{Morabito2003}. Based on the assumption that the density spectrum of irregularities follows a power-law with a Kolmogorov spectral index of $p=11/3$, an empirical relation between TEC and the spectral broadening, as proposed by \cite{Ho2002} and employed in several previous studies  \citep{Woo1977, Woo1978, Ho2002, Morabito2003, Yunqiu2015}, can be given as :

\begin{equation}
    TEC  = f \left( \frac{B_S}{c_0 } \right)^{ \left( \frac{5}{6} \right)}
    \label{eq:TEC}
\end{equation}

where $c_0 = 1.14 \times 10^{-24}$, $f$ is signal frequency in GHz, the TEC is the total electron content along the LOS path (in $m^{-2}$). The estimated TEC can then be further used to derive a rough estimate of the electron density at varying heliocentric distances using the following relation as suggested by \cite{Bird1990}:

\begin{equation}
    N_e = \frac{TEC}{r\times [ESP]}
    \label{eq:N_e}
\end{equation}
where $ESP$ is the Earth-Sun-Probe angle in radians, $N_e$ is the electron density (in m$^{-3}$), and $r$, which is the heliocentric distance, is in meters. This formula assumes a spherically symmetric coronal density and a steady state outflow.

\begin{figure}[htb]
\centering
 \includegraphics[scale=0.4]{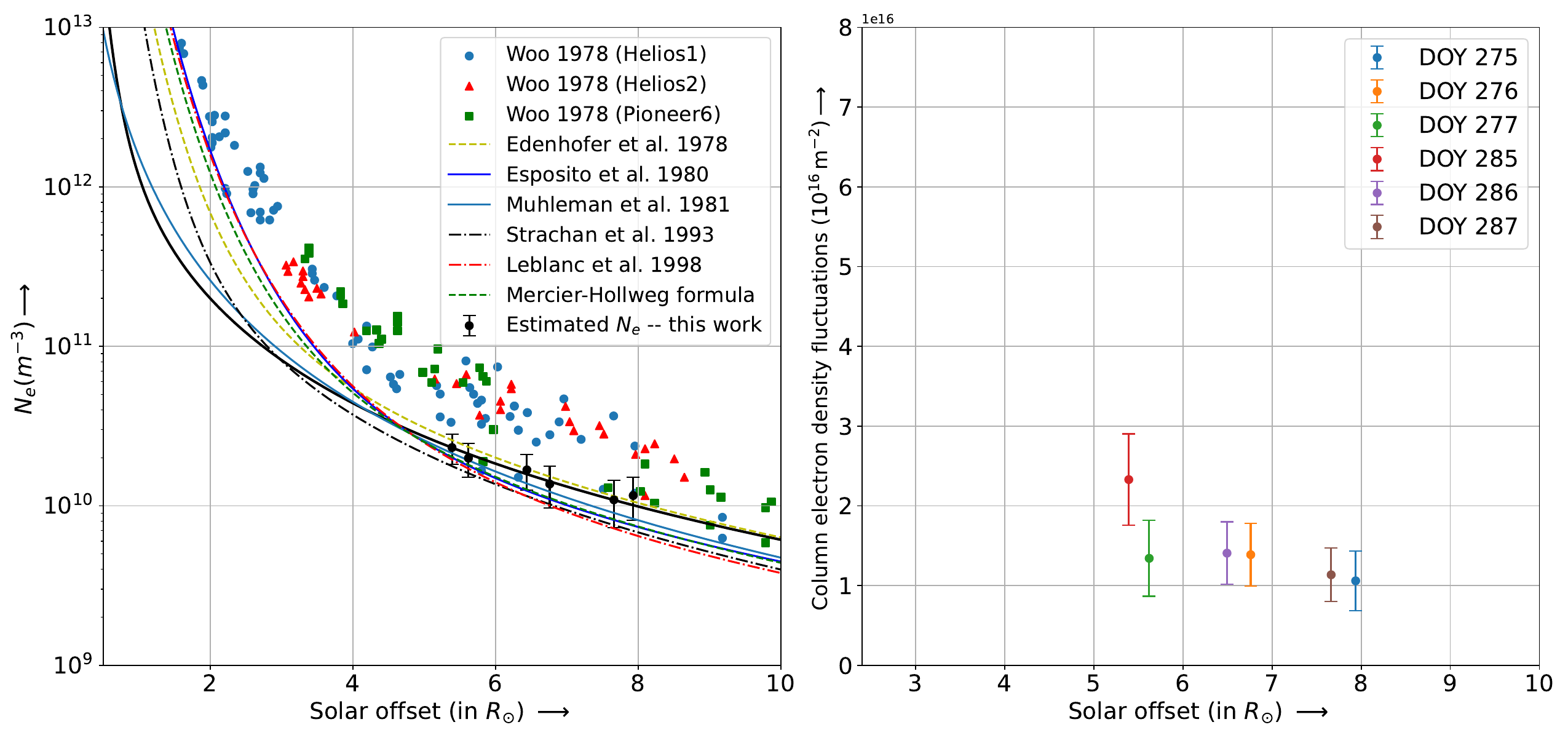}
 \caption{(a) Electron number density compared with other models present in literature (2 to 10 $R_{\odot}$). The red points represent the derived values from our study. (b) Column electron density fluctuations as measured for the region of study.}
\label{fig:densities_2}
\end{figure}

In the left panel of Fig \ref{fig:densities_2} we show our measurements for $N_e$ using MOM data and their comparison with the model $N_e$ profiles derived using different methods of observations and the density values derived using the broadening measurements from HELIOS 1$\&$2 and Pioneer 6 \citep{Woo1978}. We may note that our estimated values of $N_e$ match quite well with the model values as estimated in various studies. A general form of the equation which satisfies various observations \citep{Edenhofer1977, Esposito1980, Muhleman1981, Strachan1993, Leblanc1998, Wexler2019} can be written as :

\begin{equation}
    N_e = N_0 \times \left( \frac{A}{r^{\alpha}} + \frac{B}{r^{\beta}} + \frac{C}{r^{\gamma}} \right)
    \label{eq:equationdensity}
\end{equation}

Here $N_0, \alpha, \beta, \gamma, A, B$, and $C$ are the constant parameters having different values for different models as mentioned in the table \ref{tab:denmodel}. $r$ represents the solar-offset distance.

\setlength{\tabcolsep}{10pt} 
\renewcommand{\arraystretch}{1} 

\begin{deluxetable}{cccccc}
\tablenum{2}
\label{tab:denmodel}
\tablecaption{Electron number density $\left(N_e \right)$ models in the literature.}
\tablewidth{0pt}
\tablehead{
\colhead{Reference} & \colhead{$N_0$} & \colhead{$\frac{A}{r^{\alpha}}$} & \colhead{$\frac{B}{r^{\beta}}$} & \colhead{$\frac{C}{r^{\gamma}}$}
}
\startdata
\hline
\cite{Edenhofer1977} & $10^{12}$ & $\frac{30}{r^6}$ & $\frac{1}{r^{2.2}}$ & 0 \\
\cite{Esposito1980} & $10^{12}$ & $\frac{100}{r^6}$ & $\frac{0.5}{r^{2.1}}$ & 0 \\
\cite{Muhleman1981} & $10^{12}$ & $\frac{1.32}{r^{2.7}}$ & $\frac{0.23}{r^{2.04}}$ & 0 \\
\cite{Strachan1993} & $10^{12}$ & $\frac{15.2}{r^{6.71}}$ & $\frac{1}{r^{2.4}}$ & 0 \\
\cite{Leblanc1998} & $10^{12}$ & $\frac{80}{r^6}$ & $\frac{4.1}{r^4}$ & $\frac{0.33}{r^2}$ \\
Mercier-Hollweg formula \cite{Wexler2019} & $10^{12}$ & $\frac{65}{r^{5.94}}$ & $\frac{0.768}{r^{2.25}}$ & 0 \\
This study & $10^{12}$ & $\frac{0.296}{r^{6}}$ & $\frac{0.865}{r^{2.150}}$ & 0 \\
\enddata
\end{deluxetable}

Applying the dual power formula for electron density as a curve fit to our data, we incorporated the results as an empirical relation given as equation \ref{eq:power-law}.
\begin{equation}
    N_e(r) = A r^{-\alpha}+B r^{-\beta}
    \label{eq:power-law}
\end{equation}

A comparison of our observations with different models for electron density in the interplanetary medium is listed in the table \ref{tab:denmodel}. We may note that the decrease in the electron densities with respect to increasing helio-centric distance (Figure \ref{fig:densities_2}, left panel) matches quite well with the theoretical models \citep{Edenhofer1977, Esposito1980, Muhleman1981, Strachan1993, Leblanc1998, Wexler2019}. Additionally, applying the same approach to the broadening measurements using HELIOS 1\&2 and Pioneer-6 by \cite{Woo1978} gives similar results with MOM estimates falling at the lower edge of the distribution. A possible reason for the lower estimates could be associated with the solar activity. As measurements using MOM were made observations during the period of an extended weak solar activity period, the changes in the electron density with respect to the varying helio-centric distances are not as steep while having a comparatively lower magnitude compared to previous measurements which were made during previous solar cycles of comparatively higher solar activity \citep{Edenhofer1977, Esposito1980, Muhleman1981, Strachan1993, Leblanc1998, Wexler2019}. The solar activity was near maximum when Pioneer-6 measurements were conducted, while the measurements by HELIOS 1\&2 were during solar minimum (F10.7=150, as compared to F10.7=100 during MOM) for the Solar Cycle 20 in their highly elliptical orbits around the Sun \citep{Woo1978}.

Further, we can use the measurements of $\Delta \Omega$, as given in Equation \ref{eq:mean}, to estimate the fluctuations in the column electron densities during the period of the measurements as

\begin{equation}
    \Delta TEC = \frac{c f_{Hz}}{\kappa} \Delta \Omega
\end{equation}

where $f_{Hz}$ is the transmitted frequency in Hz, and $\kappa \approx 40.3 m^3/s^2$ \citep{Ando2015, Jain_2024}. In the right panel of Figure \ref{fig:densities_2}, we present the daily mean of the column density fluctuations as a function of increasing solar offset. As the fluctuations remain of the same order throughout the region being probed, this is a good indicator of the turbulence being diffused for the period of the experiment being a relatively quiet period. One of the measurable effects of this turbulence is angular broadening, a phenomenon caused by random phase shifts in radio wavefronts as they pass through density irregularities in the solar wind and corona.

\subsection{Estimation of Solar wind velocity using Doppler Spectral Width}

It is known that the turbulent density fluctuations in the solar wind and corona create random phase shifts in the wavefront, distorting its planar shape \citep{Hollweg1970b, Coles1989, Cairns1994}. This process, known as angular broadening, causes unresolved radio sources to appear larger and blurrier, with faster solar wind flows and stronger density variations leading to greater effects. Additionally, density fluctuations in the outer heliosphere can significantly widen the radiation angle, sometimes by 2–3 kHz, and may shift the apparent origin of the radiation. Angular broadening therefore reveals important details about the turbulence in the solar wind, the spatial distribution of density irregularities, and the scales of energy transfer, making it essential for studying magnetohydrodynamic (MHD) turbulence and solar wind acceleration \citep{Hollweg1970b, Cairns1994,Kenny2020, Tasnim2020, Wang_2024}.

Angular broadening in the radio signals can be expressed after \cite{Coles1989} as

\begin{equation}
    \theta = \frac{\frac{1}{2} r_e \lambda^2 N_e r R_{\text{SP}}}{1+R_{\text{SP}}}
    \label{eq:theta}
\end{equation}
where $r_e$ is the classical electron radius in m, $\lambda$ is the wavelength in cm, $N_e$ in $m^{-3}$, $R_{SP}$ in $AU$ and $r$ is in solar radii. Here we consider a spherically symmetric coronal density and a steady state outflow. It is worth mentioning here that \cite{Coles1989} define $\theta$ as the angular position shift of the source in radians, which is equivalent to the angular broadening.

It has been observed that angular broadening is dependent on spectral broadening, while being proportional to the plasma velocity perpendicular to the line of sight ($v_{\perp}$). As proposed by \cite{Woo1977}, the solar wind velocity as a function of angular and spectral broadening can be written as :

\begin{equation}
     v = 5.77 \frac{B_s}{\theta} \frac{R_{EP}}{k \times R_{SP}} \left( sin^2\phi + \frac{cos^2\phi}{Q^2} \right)^{-1/2}  \times \left( sin^2\gamma + Q^2cos^2\gamma \right)^{-1/2}
 \end{equation}
Here, $k$ is the spatial wave number, and $Q$ is the axial ratio of the ellipse representing the contour of equal intensity in the angular spectrum of a point source broadened by turbulence. $\gamma$ is the angle between the plane in which the angular broadening $\theta$ is measured and the magnetic field in the plane transverse to the ray path while $\phi$ is the angle between the velocity vector and the magnetic field. $Q=1$ for isotropic turbulence, and $\gamma = \phi = 0$ for the case when the solar wind is parallel to the magnetic field \citep{Woo1977,Woo1978}. This is a reasonable assumption as the solar wind is radially outwards and the magnetic field is as well at these distances \citep{Waldmeier1977}, thus reducing the equation to

\begin{equation}
\label{eq:velocity}
    v_{\perp} = \frac{5.77 \times B_S \times R_{EP}}{\theta \times k \times R_{SP}}
\end{equation}

Using the above equations \ref{eq:variance}, \ref{eq:TEC}, \ref{eq:N_e}, \ref{eq:theta} in \ref{eq:velocity}, the final reduced form of the equation for finding solar wind velocity using the spectral broadening in the received comes out to be -

\begin{equation}
\label{eq:method}
    v_{\perp} = k_0 \times \left[ \frac{[ESP]\times  R_{EP} \times (1+R_{SP})}{R_{SP}^2} \right]  \times B_S^{\frac{1}{6}} 
\end{equation}
which simplifies as 
\begin{equation}
    v_{\perp} = k_0 \left[ \frac{r\times  R_{EP} \times (1+R_{SP})^2}{R_{SP}} \right] B_S^{\frac{1}{6}}
\end{equation}
where
\begin{equation}
    k_0 = \frac{5.77 \times c_0^{5/6} \times R_{\odot} \times 10^8}{\pi \times c \times r_e} = 1.687
\end{equation}

This value of $k_0$, when used in the reduced equation for solar wind velocity (eq \ref{eq:method}), we get the solar wind velocity in the radial direction in km/s.
The distance from the spacecraft to the receiving station is indicated by $R_{EP}$, while $r$ indicates the distance from the Sun to the point of closest approach in Solar radii, $[ESP]$ is the Earth-Sun-Probe angle in radians, and $R_{SP}$ is the distance from the Sun to the probe as has been shown in figure \ref{fig:experiments}. It is important to note that while the geometry factor predominantly determines the solar wind velocities, the broadening component acts as a scaling factor and serves as a valuable indicator of turbulence in the probed region, directly influencing the observed velocity variations. The constant term of 5.77 in the calculation of $k_0$ is dependent on factors such as direction of the coronal magnetic field and angular broadening pattern, but is independent of strength of coronal turbulence \citep{Bird1982}.

\begin{figure}[htb]
\centering
\includegraphics[width=\linewidth]{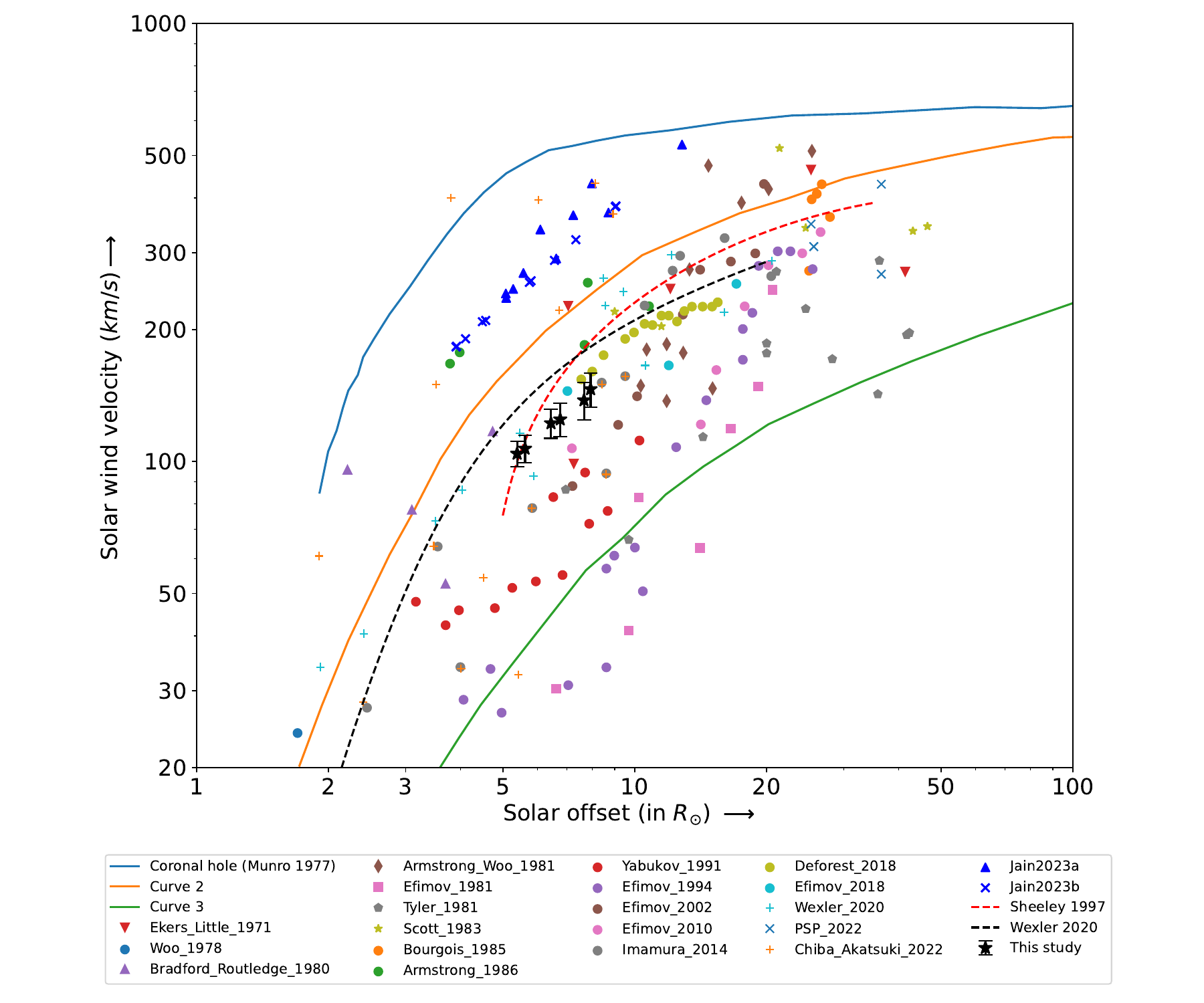}
\caption{Our results compared against the solar wind measurements done across a period of more than 50 years, using a variety of methods from in-situ to remote sensing observations. The velocity profile of the solar wind within a polar coronal hole having a cross-sectional area denoted by $A(r)$ is depicted by Curve 1. This profile is derived from observations made using Skylab's white-light instruments \citep{Munro1977}. Additional models for the volume velocity of the solar wind with and without contributions from MHD waves are represented by Curves 2 and 3 \citep{Esser1986}. \label{fig:history}}
\end{figure}

\subsection{Comparison of estimated Solar wind velocity with observations from different missions}

In Figure \ref{fig:history}, we compare results from our study against Solar wind velocity measurements done employing different methods over a span almost 50 years using different missions during the years 1971-2023.  These studies include IPS studies of Goldstone Deepspace Tracking Station, California \citep{Ekers1971}, HELIOS 1$\&$2 and Pioneer 10$\&$11 \citep{Woo1978}, Voyager 2 \citep{Bradford1980}, HELIOS 1 \citep{Armstrong1981}, Venera 10 \citep{Efimov1981}, Viking and Mariner 10 \citep{Tyler1981}, Multistation IPS \citep{Scott1983}, EISCAT antenna \citep{Bourgois1985}, VLA with radio source 3C279 \citep{Armstrong1986}, Venera 15-16 \citep{Yakubov1991,Efimov1994}, Galileo \citep{Efimov2002}, NOZOMI \citep{Efimov2010}, Akatsuki \citep{Imamura2014, Wexler2020, Chiba2022, Jain_2023, Jain_2024}, STEREO \citep{DeForest2018}, Galileo, SOHO, Mars Express, Venera 15-16 \citep{Efimov2018}, and the Parker Solar Probe (PSP). In figure \ref{fig:history} the velocity profile of the solar wind within a polar coronal hole having a cross-sectional area denoted by $A(r)$ derived from observations made using Skylab's white-light instruments has been shown in a solid blue curve which demonstrates a pronounced acceleration within the range from $2$ to $5 R_{\odot}$, and this behavior is attributed to the conservation of mass flow, beyond which it is based on a solar wind model for high-speed flows \citep{Munro1977, Leer1982}. Additional models for the volume velocity of the solar wind with and without contributions from MHD waves are represented by Curves 2 and 3 in figure \ref{fig:history} \citep{Esser1986}. Results from \cite{Sheeley1997, Wexler2019, Wexler2020, Wexler_2020} have also been added to compare the theoretical curves 2 and 3, and to put our measurements in this study in perspective.

Other studies from which results have been included are  \cite{Bourgois1985, Armstrong1986, Armstrong1972, Jokipii1973} and \cite{Bradford1980}, where the occulted sources are located above the solar north pole, which minimizes the effect of line-of-sight.

\subsection{Estimation of errors in Solar wind velocity measurements}

It is clear from equation \ref{eq:method}, which we have used for estimating solar wind velocity in this study that the error in the velocity estimation depends on only a single parameter: spectral broadening of the signal, as the constant $k_0$ and the geometry are taken to be free of error. For the calculation of error, as the geometrical term is derived using the NASA SPICE toolkit, it is assumed to be free of error. Similarly, the constant $k_0$ is also with no errors, as it is the resultant value of all the constants being used in the derivation. Thus, the only factor which contributes to the error in measurement of solar wind velocities is the value of spectral broadening in the signal. From this factor, apart from the Doppler component due to the relative motion of the probe and the receiving ground station which can be easily removed, the broadening in the Doppler spectrum which gets introduced due to the rate of change of the LOS velocity can be removed as shown in fig \ref{eq:LOS_doppler}, the results of which have been shown in fig \ref{fig:LOS_mitigation} \citep{Tripathi2022b}. Further, doing a moving average removes the extreme outliers, and for the Allan variance of the oscillator of $10^{-11}$ stability, the error due to the instability in the generated signal for the frequency is $10^{-11}\times f = 0.0229$ Hz \citep{Tripathi2022b}.

According to \citep{Taylor1997}, the fractional uncertainty in a measurement $y= x^{\frac{1}{6}}$ can be given as

\begin{equation}
    \frac{\delta y}{|y|} = \frac{1}{6} \frac{\delta x}{|x|}
    \label{eq:taylor}
\end{equation}
Thus, since we are using the $1/6$th power of the spectral broadening value ($B_s$), the error also scales accordingly as $1/6$th of the error in spectral broadening measurement which has been used as a measure of error in the solar wind velocities derived in this study as per equation \ref{eq:taylor}. Taking the sum of these two values gives us the total error in $B_s ^{1/6}$.

Table \ref{tab:combined} gives the values of solar wind velocities derived using equation \ref{eq:method} for the Radio-occultation Observations conducted in October 2021, along with the error estimates in the measurements for the same. Fig \ref{fig:history} shows our measurements in black points with error bars.

\begin{deluxetable*}{cccccc}
\tablenum{5}
\tablecaption{Combined Solar Wind Velocities and Errors measured by MOM (In km/s)\label{tab:combined}}
\tablewidth{0pt}
\tablehead{
\colhead{Date} & \colhead{Solar Offset (R)} & \colhead{MOM $v_{\perp}$ (km/s)} & \colhead{Error (km/s)} & \colhead{Error (\%)}
}
\startdata
2nd Oct 2021 & 7.93 & 146.35 &$\pm$ 13.11 &$\pm$ 8.95 \\
3rd Oct 2021 & 6.76 & 124.83 & $\pm$10.96 &$\pm$ 8.78 \\
4th Oct 2021 & 5.62 & 107.07 & $\pm$7.78 &$\pm$ 7.27 \\
12th Oct 2021 & 5.39 & 104.27 & $\pm$6.85 &$\pm$ 6.57 \\
13th Oct 2021 & 6.49 & 122.37 &$\pm$ 9.35 &$\pm$ 7.64 \\
14th Oct 2021 & 7.66 & 138.07 & $\pm$13.46 &$\pm$ 9.75 \\
\enddata
\end{deluxetable*}

\section{Concluding remarks}
\label{sec:ANSWER}

We used the Radio occultation measurements conducted by the Indian MOM spacecraft to gain insight into the behavior of the Solar wind in the middle and outer coronal regions in this work. Using spectral broadening of the received signals, we obtained velocities of solar wind between $100-150$ km/s for the heliocentric distances 5 - 8 $R_\odot$ from October 2 to October 14, 2021, the period during which the experiments were conducted. The results were derived using equation \ref{eq:method}, which we propose as a general equation for the derivation of solar wind velocities using the spectral broadening in the signal and the associated geometrical component. The velocities of solar wind, derived in the presented radio occultation study, agree with previously reported values (ref to fig \ref{fig:history}), and show the acceleration of the solar wind in the Middle Solar Coronal region. Though the variation of electron density with heliocentric distance was found to be relatively flat and exhibited a comparatively lower magnitude than earlier measurements, the derived density profile matches with the model values (Table \ref{tab:denmodel}). The difference may be attributed to the prolonged weak solar activity during the MOM observations, in contrast to prior studies conducted during periods of higher solar activity in earlier solar cycles. While the heliocentric distances from the Sun for which observations in this study were made were very limited, the results demonstrate the reliability of the approach used and provide useful insights into the solar wind behavior during this period.

\section*{Acknowledgments}
The valuable inputs provided by the reviewer, Dr. David Wexler, and the suggested references were deeply valued, leading to significant enhancements in the article's quality.
We express our gratitude to the dedicated team at the Indian Deep Space Network in Bangalore, India, for diligently monitoring the radio signals transmitted by MOM using the IDSN32 antenna. Additionally, we extend our special appreciation to the MOM mission team at UR Rao Satellite Center in Bangalore for granting us the privilege to track the signals from MOM during the solar conjunction. The authors greatly appreciate the help from Richa N. Jain, Ajay Potdar, and Soumyaneal Banerjee from InSWIM Lab at SPL. Help provided by R. S. Simi in providing the data is also gratefully acknowledged. Author KA has received a research fellowship (PMRF-2103356) from the Prime Minister's Research Fellowship (PMRF) scheme, Ministry of Education, Government of India. The authors KA and AD acknowledge the use of facilities procured through the funding via the Department of Science and Technology, Government of India sponsored DST-FIST grant no. SR/FST/PSII/2021/162 (C) awarded to the DAASE, IIT Indore.

\section*{Data Availability}
Solar occultation data from MOM, which can be obtained from the Indian Space Science Data Center (ISSDC) was used for the experiment. Position maps were created using data from the Indian Space Science Data Center (ISSDC) and the NASA SPICE toolkit. 

\bibliography{sample631}
\end{document}